\documentstyle[epsfig]{article}

\newcounter{eqnletter}[equation]
\catcode`\@=11 \@addtoreset{equation}{section} \catcode`\@=12

\begin{document}
 \begin{centering}
{\large \bf Probability density of determinants of random matrices.}\\
\vskip1cm
Giovanni M. Cicuta\\ 
Dipartimento di Fisica, Univ. di Parma, Viale delle Scienze 7A, \\
43100 Parma, Italy, and\\ Gruppo collegato di Parma dell'INFN, Sez. di Milano\\
e-mail : cicuta@fis.unipr.it
\vskip .5cm
Madan Lal Mehta \\
CEA/Saclay, Service de Physique Th\'eorique,\\
F-91191 Gif-sur-Yvette Cedex, France\\
e-mail : mehta@spht.saclay.cea.fr

\end{centering}
\vskip 1cm
\noindent {\bf Abstract.}  In this brief note the probability density of a 
random real, complex and quaternion determinant is rederived using the singular
values. The behaviour of suitably rescaled random determinants is studied in the limit of infinite order of the matrices.
\bigskip

\noindent\section{\bf Introduction and results.} 
We consider $n\times n$ matrices whose elements are either real, or complex or 
quaternions (in what follows, the four components of the quaternions will
always be real); 
the real parameters entering these elements are independent
gaussian random variables with mean zero and the same variance.
 The number of real
parameters needed to characterize an $n\times n$ matrix is thus $\beta n^2$,
where $\beta$ is 1, 2 or 4 according as the matrix elements are real, complex or
quaternions. We will derive the probability density of their determinant.

The determinant of random real matrices is an old subject \cite{kull}, that of random
complex matrices and of random hermitian complex matrices was studied some
time back \cite{mehnor} , that of random quaternion matrices presents some
peculiar features due to the non-commutative multiplication, as we will see
below while the case of real symmetric matrices has been settled recently for odd $n$
\cite{dela}. 

The method we will use here is to start with the joint probability
density of the singular values rather than that of the eigenvalues. As the absolute
value of the determinant is the product of all the singular values, we can find its
probability density by calculating its Mellin transform. This gives new proofs of the
known results for random real and complex matrices and of a partial result for
quaternion matrices. 

For quaternions, multiplication being not commutative, it is not possible to define
a determinant having the usual three properties \cite{meh1}; namely, (i) $\det\, A=0$ if
and only if $Ax=0$ has a
non-zero solution $x\neq 0$, (ii) $\det(AB)=\det\, A.\det\, B$,
(iii) $\det\, A$ is multi-linear in the rows of $A$. So the definition of a determinant
varies according to which of the property or properties one wants to keep. We
will adopt the following definition due to Dieudonn\'e or Artin \cite{art}.

Any matrix $A$ is either singular (i.e. $Ax=yA=0$ have non-zero solutions) or
has an inverse (i.e. $AB=BA=I$) \cite{meh1}. If $A$ is singular, define $\det\, A=0$. If
$A$ has
an inverse, define $\det\, A$ by recurrence on $n$ as follows. If $n=1$, define
$\det\, A=\vert a_{11}\vert$, where $\vert x\vert$ means the norm of (the
quaternion) $x$. If $n>1$, then let $A_{ij}$ be the $(n-1)\times(n-1)$ 
matrix obtained by removing the $i$-th row and the $j$-th column of $A$. The
matrix elements of $B$, the inverse of $A$, are written as $b_{ij}$. Not all $b_{ij}$
are zero. One shows \cite{art} that whenever $b_{ij}=0$, $\det\, A_{ji}=0$, and whenever
$b_{ij}\ne 0$, $\det\, A_{ji}\ne 0$ and $\vert b_{ij}^{-1} \det\, A_{ji}\vert$ is
independent of $i$ or $j$. One then defines $\det\, A = \vert b_{ij}^{-1} \det\,
A_{ji}\vert$. Thus for a quaternion matrix $A$, $\det\,A$ is a non-negative real
number. (For real or complex $A$ this definition also gives a non-negative real
number, the absolute value of the usual ordinary determinant.) 
Note that this determinant is not linear in the rows of $A$, but has the other two
properties \cite{art}. Also that a quaternion matrix $A$ may be singular while its
transpose has an inverse \cite{meh1}. 

The eigenvalues and eigenvectors of a matrix $A$ are defined as the solutions of
$ A\varphi=\varphi x$, where $\varphi$ is an $n\times 1$ matrix and $x$ is a
number.  For a real or complex $A$ one can eliminate $\varphi$ to get
$\det(A-xI)=0$, where $I$ is the unit matrix. For quaternion $A$, if $x$ is an
eigenvalue with the eigenvector $\varphi$ and $\mu$ any constant quaternion,
then $\mu^{-1}x\mu$ is an eigenvalue with the eigenvector $\varphi\mu$. Thus
$x$ and $\mu^{-1}x\mu$ are not essentially distinct as eigenvalues. It is
not evident that an $n\times n$ quaternion matrix should have $n$ (quaternion) 
eigenvalues, but it has \cite{meh}.  One can actually put them in correspondance
with complex numbers \cite{6}. Here we will only note   that the norm of the product of eigenvalues gives the determinant defined above. 

If all the eigenvalues of $A$ are essentially distinct, then one can diagonalise
$A$ by a non-singular matrix. To make things clearer, we give an example:  
\begin{eqnarray} 
\left[ \matrix{1&e_2\cr e_1&e_3\cr}\right] \left[ \matrix{1&1\cr e_2&e_1\cr}\right]
& = & \left[ \matrix{1&1\cr e_2&e_1\cr}\right]  \left[ \matrix{0&0\cr 0&1-e_3\cr}\right], 
 \label{a.1}
\end{eqnarray}
\begin{eqnarray} 
\left[ \matrix{1&e_1\cr e_2&e_3\cr}\right] 
\left[ \matrix{1&1\cr a & b \cr}\right] & =& 
\left[ \matrix{1&1\cr a & b \cr}\right]
\left[ \matrix{ x_1 & 0 \cr 0 & x_2 \cr}\right], 
 \label{a.2}
\end{eqnarray}
with 
 \begin{eqnarray} 
a = {1\over 2}(1-\sqrt 3)(e_1-e_2)\quad ,  \quad 
b = {1\over 2}(1+\sqrt 3)(e_1-e_2),
 \label{a.3}
\end{eqnarray}
 \begin{eqnarray} 
x_1 = {1\over 2}(1+\sqrt 3) - {1\over 2}(1-\sqrt 3)e_3 \quad , \quad
x_2 = {1\over 2}(1-\sqrt 3) - {1\over 2}(1+\sqrt 3)e_3, \qquad
 \label{a.4}
\end{eqnarray}
showing that the eigenvalues of $\left[\matrix{1&e_2\cr e_1&e_3}\right]$ are $0$
and $1-e_3$, while those of its transpose $\left[\matrix{1&e_1\cr e_2&e_3}\right]$
are $x_1$ and $x_2$. Their determinants are respectively $0$ and $2$. \\

If all the eigenvalues $x_i$ are real and positive
(respectively, real and non-negative), one says that $A$ is positive definite
(respectively, positive semi-definite). Denote by $A^{\dag}$ the transpose,
hermitian conjugate or the dual of $A$ according as $A$ is real, complex or
quaternion. For any matrix $A$, the product  $AA^{\dag}$ (or $A^{\dag} A$) is
positive semi-definite, its eigenvalues are real and non-negative. The positive
square roots of the eigenvalues of $AA^{\dag}$ (or of $A^{\dag} A$, they are the
same) are known as the singular values \cite{7} of $A$. The eigenvalues and
singular values of $A$ have, in general,  nothing in common, except that 
\begin{eqnarray} 
 \prod_{i=1}^n \lambda_i^2 = \det(AA^{\dag}) = \det\, A.\det\, A^{\dag} 
= \vert \det\, A\vert^2 = \prod_{i=1}^n\vert x_i\vert^2,\label{a.5}
\end{eqnarray}
where $\lambda_i$ are the singular values and $x_i$ are the eigenvalues of $A$. 

In section 2 we start with the joint probability density of the singular values and
calculate the Mellin transform of the probability density $p(\vert y\vert)$ of the
(absolute value) of the determinant $y$ of a random matrix $A$. From symmetry, 
when $A$ is real, $y$ is 
real and $p(y)$ is even in $y$; when $A$ is complex, $y$ is complex and $p(y)$
depends only on $\vert y\vert$. When $A$ is quaternion, $y$ is  by definition real
and non-negative. One can therefore recover $p(y)$ from $p(\vert y\vert)$ when
$A$ is real or complex.  
Our  results, confirming those in the known cases, are as follows. 
 \begin{eqnarray} 
p_1(y) &=& \prod_{j=1}^n [\Gamma(j/2)]^{-1} G^{n,0}_{0,n}\left(y^2 \left| \, 0,
{1\over 2},  {2\over 2}, {3\over 2},...,{n-1\over 2} \right. \right), \quad  y \; {\rm real},  
 \label{a.6}\\
p_2(y) & =& {1\over \pi} \prod_{j=1}^n [\Gamma(j)]^{-1} G^{n,0}_{0,n}\left(\vert
y\vert^2 \left| \right. 0,1,2,...,n-1 \right),  \quad y\; {\rm complex},    \label{a.7}\\
p_4(y) & =& 2 \prod_{j=1}^n [\Gamma(2j)]^{-1}  G^{n,0}_{0,n} \left(y^2\left| \, {3\over
2}, {7\over 2}, {11\over 2},...,2n-{1\over 2} \right. \right), \quad y\; {\rm real\
non\ negative}.  \nonumber \\  \label{a.8}
\end{eqnarray}
Here $ G^{n,0}_{0,n} $ is a Meijer G-function. In the above results the gaussian  probability distribution $P(A)$ for the matrix $A$ was taken $P(A) \propto e^{-a \, {\rm tr} A^\dag A}$ , with $a=1$. Next we show that the probability density of the random variable $y= [\det A^\dag A]^{1/n}$ converges to  $\delta (
y- 1/e)$ in the large $n$ limit , with $P(A) \propto e^{-a \, {\rm tr} A^\dag A}$ and $a=\beta \, n/2$. In section 3 we study the large $n$ limit for a non-gaussian random complex matrix and show that the random variable $y= [\det A^\dag A]^{1/n}$ converges in the large $n$ limit to a constant, whose value, depending on the parameters in the non-gaussian probability distribution, is different in the two phases of the model. Finally for a hermitian complex random matrix $H$ with probability density
$P(H) \propto e^{-n \, {\rm tr} H^2 }$ , we show that in the large $n$ limit $|\det H|^{1/n}$ tends to the constant $1/ \sqrt{2 e}$.\\

Some of these results are probably known to some experts, since analogous results appear in the literature \cite{past}.
 
\bigskip

\noindent\section  {\bf Gaussian matrices.} 
The joint probability density of the singular values can conveniently be derived in
two steps from the two observations \cite{8},\cite{and}

(i) Any matrix $A$ can almost uniquely be written as 
$U\Lambda V$  where
$\Lambda$ is a diagonal matrix with real non-negative diagonal elements, while
$U$ and $V$ are real orthogonal, complex unitary or quaternion symplectic
matrices according as $A$ is
real, complex or quaternion; ``almost uniquely" refering to the fact that either $U$
or $V$ is undetermined up to multiplication by a diagonal matrix.  

(ii) Any positive semi-definite matrix $H = AA^{\dag}$ can be
written uniquely as $H=TT^{\dag}$, where $T$ is a triangular matrix with real
non-negative diagonal elements.

As a result the gaussian joint probability density $\exp(-a \,{\rm tr}\, AA^{\dag})$ for
the matrix elements of $A$ gets transformed to  
\begin{eqnarray} 
 F(\Lambda) \equiv F(\lambda_1, ... , \lambda_n) = {\rm const.}\,
\exp\left(-a \sum_{j=1}^n \lambda_j^2\right)
\vert\Delta(\lambda^2)\vert^\beta \prod_{j=1}^n \lambda_j^{\beta-1} \label{a.12}
\end{eqnarray}
where $\lambda_1,..,\lambda_n$ are the singular values of $A$, 
$\Delta$ is the product of differences 
\begin{eqnarray} 
 \Delta(\lambda^2) = \prod_{1\le j<k\le n}(\lambda_k^2-\lambda_j^2)\label{a.14}
\end{eqnarray}
and $\beta=1$, 2 or 4 according as $A$ is real, complex or quaternion. 

The Mellin transform of the product of the $\lambda$'s is 
\begin{eqnarray} 
{\cal M}_n(s) & = &{\rm const.}\, \int_0^\infty \eta^{s-1}
\delta(\eta-\lambda_1...\lambda_n)\,  
F(\Lambda) \,d\lambda_1 ... d\lambda_n  \,d\eta  \nonumber \\
& = & {\rm const.}\, \int_0^\infty 
\exp\left(-a \sum_{j=1}^n \lambda_j^2\right)
\vert\Delta(\lambda^2)\vert^\beta \prod_{j=1}^n \lambda_j^{\beta+s-2} 
d\lambda_j \nonumber \\
& = & {\rm const.}\, \int_0^\infty 
\exp\left(-a \sum_{j=1}^n t_j\right)
\vert\Delta(t)\vert^\beta \prod_{j=1}^n t_j^{(\beta+s-3)/2} 
dt_j \nonumber \\
& = & a^{-n(s-1)/2}\,\prod_{j=1}^n \left[\frac{\Gamma\left({s-1\over 2}+{j\beta\over
2}\right)}{\Gamma\left({j\beta\over 2}\right)}\right]   \label{a.15}
\end{eqnarray}
In the last line we have used a result derived from Selberg's integral \cite{10}.
The constant, independent of $s$, is fixed from the requirement that ${\cal
M}_n(1)=1$. 
\bigskip 

The inverse Mellin transform of the expression  (\ref{a.15}) is a Meijer G-function \cite{erd}

\begin{eqnarray} 
p_\beta (\vert y\vert) = 2\, a^{n/2}\prod_{j=1}^n \left[\Gamma(j\beta/2)\right]^{-1}
G^{n,0}_{0,n}\left(a^n |y|^2 \left| {\beta-1\over 2},\right. {2\beta-1\over 2}, ..., {n\beta-1\over
2}\right) 
\label{a.16}
\end{eqnarray}
When $\beta=1$, the matrix $A$ is real, its determinant $y$ is real, from symmetry
the probability density $p_1(y)$ is an even function of $y$ and we have 
\begin{eqnarray} 
 p_1(y) = {1\over 2} p_1(\vert y\vert) 
\label{a.17}
\end{eqnarray}
giving equation (1.6) with $a=1$ . When $\beta=2$, $A$ is complex, $y$ is complex, from
symmetry $p_2(y)$ depends only on the absolute value $\vert y\vert$ of $y$, and
one has for $a=1$ ,
\begin{eqnarray} 
 p_2(y) & =& {1\over 2\pi\vert y\vert}p_2(\vert y\vert)   \nonumber \\
& =& {1\over\pi} \prod_{j=1}^n
[\Gamma((j)]^{-1} {1\over \vert y\vert}G^{n,0}_{0,n}\left(|y|^2\left| {1\over 2},\right.
{3\over  2}, ..., n-{1\over 2}\right)  \nonumber \\
& =& {1\over\pi}\prod_{j=1}^n[\Gamma(j)]^{-1}G^{n,0}_{0,n}(|y|^2\left| 0,1,...,n-1) 
\right.\label{a.18}
\end{eqnarray}
which is equation (1.7). 
When $\beta=4$, $A$ is quaternion, $y$ is, by definition, real positive, and $p_4(y) =
p_4(\vert y\vert)$, giving equation (1.8). 

In reference 2, appendix A.5, we somewhat conventionally mapped the
(quaternion) eigenvalues on to the essentially equal eigenvalues having the
scalar part and only one other component at most, accounting for a factor $y^2$
in the probability density. Moreover, equation (A.43) there has a misprint,
$\Gamma((s+2j+1)/2)$ there should read $\Gamma((s/2)+2j+1)$. Thus equation
(1.8) tallies with equation (A.43) of reference 2. \\

We now evaluate the large $n$ behaviour of the moments $<y^k>$  of the
random variable $y= [\det A^\dag A ]^{1/n}$ which show that it
converges to a  constant in the $n \to \infty$ limit. Next the same result is obtained 
 by the saddle point method. In the study of large $n$, the proper choice of 
the parameter $a$ for the gaussian ensembles  is $a=\beta \, n/2$. Then
 eq.(\ref{a.15})  implies

\begin{eqnarray}
< \Bigg(\det A^\dag A \Bigg)^k> =\left(\frac{\beta \, n}{2}\right)^{-n k}
\prod_{j=1}^{n}\left[\frac {\Gamma(k+\frac{j \beta}{2})}{\Gamma(\frac{j \beta}{2})}\right]
 \label{b.3}
\end{eqnarray}
that is
\begin{eqnarray}
&&\log< \Bigg(\det A^\dag A \Bigg)^k> =-n k\, \log \frac{\beta \, n}{2}
+\sum_{j=1}^{n}\log \frac{\Gamma(k+\frac{j \beta}{2})}{\Gamma(\frac{j \beta}{2})}
 \nonumber\\
&\approx &-n k \,\log\frac{\beta \, n}{2} +n\int_0^1 dx \log
\frac{\Gamma(k+\frac{\beta}{2}+
\frac{n x \beta}{2})}{\Gamma(  \frac{\beta}{2}+  \frac{n x \beta}{2} )}
+O(\log \, n)  \nonumber \\
&\approx &
 -n k +O(\log \,n)
\label{b.4}
\end{eqnarray}
where the Euler-Maclaurin formula has been used to estimate the large $n$ asymptotics.
When $x$ is near $0$, the integrand is a constant and its contribution is negligible. When $x$ is not small, one can ignore other  terms  compared to $nx\beta/2$.

 Replacing $k \to k/n$ 
in eqs.(\ref{b.3}),\,(\ref{b.4}), 
one gets
\begin{eqnarray} 
\lim_{n \to \infty}\log< \Bigg(\det A^\dag A \Bigg)^{k/n }
>  = -k
\label{b.6}
\end{eqnarray}
From the knowledge of all the moments (\ref{b.6}),  we conclude
 that the random variable $y=(\det A^\dag A )^{1/n}$ converges in the
large $n$ limit  to the constant $1/e$.\\

It is convenient to evaluate the above large $n$ limit  also
 by a saddle point approximation because this is easy to generalize
  to different probability distributions. Let us recall the
  asymptotic density of squared singular values (see \cite{cmmr}
eq.(10) after setting $L=1$, $m^2=1$, $g=0$, hence $A=0$, $B=4$ ) 
\begin{eqnarray}
\rho(t)=\lim_{n \to \infty} \frac{1}{n}\sum_j <\delta(t-t_j)>=
\frac{1}{2 \pi} \sqrt{\frac {4-t}{t}} \quad , \quad 0 < t \leq 4
\label{b.7}
\end{eqnarray}
Then it is easy to evaluate
\begin{eqnarray}
 \lim_{n \to \infty} \!\!&\!\!&\!\!\! \frac{1}{n} < \log \Bigg(\det A^\dag A \Bigg)^k> =
\lim_{n \to \infty}\frac{1}{n}  < k \sum_{j=1}^n \log \,t_j >
\nonumber \\
&=& k \int_0^4 \log t \, \rho(t) dt = - k
\label{b.8}
\end{eqnarray}
confirming eq.(\ref{b.6}).

\section { Large $n$ for non-gaussian complex matrices. }  Let us now consider
an example of non-gaussian probability distribution such that in the large $n$ limit two different spectral densities for the singular values exist. For simplicity, we
consider
the  ensemble of $n \times n$ complex matrices $A$ with
the non-gaussian probability distribution
\begin{eqnarray}
P(A)  \propto e^{-n (a \, {\rm tr} A^\dag A +2 b \,{\rm tr}
A^\dag A A^\dag A)} \quad , \quad b>0 \quad , \quad a \; {\rm
real} \label{b.9}
\end{eqnarray}
The analogous ensemble of real matrices would need only trivial changes.

Again the evaluation of all the moments of $y= [\det A^\dag A]^{1/n} $
 may be performed in terms
of $t_j$ , the squared singular values of the matrix $A$
\begin{eqnarray}
< \Bigg(\det A^\dag A \Bigg)^k> = \frac{\int..\int_0^\infty
\Delta^2(t) \prod_{j=1}^n t_j^k \,e^{-n( a t_j+2 b t^2_j)}\, dt_j}
{\int..\int_0^\infty \Delta^2(t) \prod_{j=1}^n \,e^{-n ( a t_j+2 b
t^2_j) }\, dt_j}
 \label{b.10}
\end{eqnarray}
Since $ \prod_{j=1}^n (t_j)^{k/n} = \exp \left(\frac{k}{n} \sum_1^n \log
t_j \right)$ , the large $n$ limit for all the moments $<y^k>$ are easily
evaluated by the saddle point approximation
\begin{eqnarray}
\lim_{n \to \infty} <y^k>= \lim_{n \to \infty}< \exp \left(\frac{k}{n}
\sum_1^n \log t_j \right)> = \exp \left(k\int dt \, \rho(t)\, \log \,t \right)
 \label{b.11}
\end{eqnarray}
where $\rho(t)$ is the solution of the saddle point equation
\begin{eqnarray}
\frac{a}{2} +2 b t = {\mathbf{-}\!\!\!\!\!\!\int}
 \frac{ \rho(y)}{t-y} dy
 \label{b.12}
\end{eqnarray}
The solution of eq.(\ref{b.12}) has two different forms, $\rho_1(t)$ , $\rho_2(t)$ 
(see \cite{cmmr} eq.(10) after setting $L=1$, $m^2=a$, $g=b$ ), depending on the values
of the real number $a/\sqrt{b}$ being larger or smaller than the
critical value $a/\sqrt{b}=-4$. At the critical value,
 $\rho_1(t)=\rho_2(t)$.
\begin{eqnarray}
\rho_1(t)&=& \frac{1}{\pi} \sqrt{ \frac{ C-t}{t}} \,\left[ 2b t +
\frac{ 2a +\sqrt{a^2+48 b}}{6}\right] \quad ,\qquad 0< t \leq C
\quad , \nonumber
\\ C &=&\frac{ \sqrt{a^2+48 b}-a}{6b} \quad , \qquad a/\sqrt{b}\geq
-4 \quad ;
 \label{b.13}
\end{eqnarray}
The definite integrals related to eq.(\ref{b.11}) for the spectral
function $\rho_1(t)$ are known in closed form and 
\begin{eqnarray}
\int_0^C \rho_1(t) \,\log t \,  dt = \log \frac{C}{4}
-\frac{1}{2}-\frac{a C}{8} \label{b.15}
\end{eqnarray}
Therefore in the \lq \lq perturbative phase", that is for $a/\sqrt{b}\geq
-4 $ , the random variable $y$ converges in the large $n$ limit to      
a  constant
\begin{eqnarray}
&& \lim_{n \to \infty} [ \det A^\dag A]^{1/n} = \frac{C}{4}\;
e^{-\frac{1}{2}-\frac{a C}{8}} 
 \label{b.16}
\end{eqnarray}

For   $a/ \sqrt{b} \leq -4$  we have
\begin{eqnarray}
&& \rho_2(t)= \frac{2 b}{\pi}\sqrt{(B-t)(t-A)} \quad , \quad
A\leq t \leq B \quad , \nonumber \\ && A+B=-\frac{a}{2 b} \quad ,
\quad B-A= \frac{2}{\sqrt{b}} \quad ,
 \qquad a/\sqrt{b}\leq -4
 \label{b.14}
\end{eqnarray}
The definite integral related to eq.(\ref{b.11}) for the
spectral function $\rho_2(t)$ may still be
evaluated
\begin{eqnarray}
I(a, b)&=&\int_A^B \rho_2(t) \,\log t \, dt =\frac{2 b}{\pi}\int_A^B\sqrt{(B-t)(t-A)} \, \log t  \, dt  \nonumber \\
&=&\frac{b}{8}(\sqrt{B}-\sqrt{A})^4+\frac{b}{2}(B-A)^2 \log\left(\frac{\sqrt{B}+\sqrt{A}}{2}\right) 
\label{b.15}
\end{eqnarray}
One may still conclude that also in the \lq \lq non-perturbative phase",
that is for $a/\sqrt{b}\leq -4 $ , the random variable $y$
converges in the large $n$ limit to a  constant
\begin{eqnarray}
\lim_{n \to \infty} [ \det A^\dag A]^{1/n} = e^{I(a, b)} \quad ,
  \qquad a/\sqrt{b}\leq -4
 \label{b.19}
\end{eqnarray}

We remark that other functions of the determinant, such as
$w=\frac{1}{n^{n^2}}\,[\det A^\dag A ]^{n},$
may not have a finite limiting probability distribution, in the large $n$ limit, 
though
their large $n$ behaviour may still be evaluated.
\bigskip

{\bf  Hermitian matrices. } The large $n$ limit of the absolute value of the determinant of gaussian hermitian matrices may be evaluated from the exact finite $n$ moments given in \cite{mehnor}. Let us consider an ensemble of $n \times n$ hermitian matrices $H$ with probability distribution $P(H) \propto e^{-n \,{\rm tr} \, H^2}$ and the random variable $y$
\begin{eqnarray}
y= | \det \, H \,|^{1/n}\, .
 \label{b.23}
\end{eqnarray}
 Let $x_j$ , $j=1,..,n$ be the eigenvalues of the matrix $H$, then
\begin{eqnarray}
<y^k>=\frac{\int..\int_{-\infty}^\infty
\Delta^2(x) \prod_{j=1}^n |x_j|^{k/n} \,e^{-n  x^2_j}\, dx_j}
{\int..\int_{-\infty}^\infty \Delta^2(x) \prod_{j=1}^n \,e^{-n  x^2_j }\, dx_j}
 \label{b.24}
\end{eqnarray}
These moments are known from \cite{mehnor}. 
\begin{eqnarray}
<y^k>=n^{- k/2} \, \prod_{j=1}^n \frac{\Gamma(\frac{1}{2}+\frac{k}{2n}+b_j^+)}
{\Gamma (\frac{1}{2}+b_j^+)} \quad , \quad b_j^+= \left[ \frac{j}{2} \right]
 \label{b.25}
\end{eqnarray}
 \begin{eqnarray}
\log <y^k> \approx -\frac{k}{2} \log ( 2 \, e) +O \left( \frac{\log n}{n} \right) .
 \label{b.26}
\end{eqnarray}
Thus in the large $n$ limit $ |\det \, H |^{1/n}$ tends to the constant $1/ \sqrt{2 e}$.

\vskip 1cm

{\bf Acknowledgments.}\\

One of the authors (M.L.M.) is thankful to the Dipartimento di Fisica, Universit\'a di Parma, 
for the hospitality where this work was done. We are thankful to L. Molinari for help in evaluating the integral (\ref{b.15}).

\end{document}